\documentclass[prb,twocolumn,amssymb,showpacs]{revtex4}
\usepackage{graphicx}
\usepackage{amsmath}

\begin{document}

\title{Ferromagnetic -spin glass transition induced by pressure in Gd$_2$Mo$_2$O$_7$}
\author{I. Mirebeau$^1$, A. Apetrei$^1$, I. Goncharenko$^1$,
D. Andreica$^{2,3}$,
 P. Bonville$^4$, J. P. Sanchez$^5$, A.
Amato$^2$,  E. Suard$^6$ and W. A. Crichton$^7$
\\
}
\address{
$^1$Laboratoire L\'eon Brillouin, CEA-CNRS, CE-Saclay, 91191
Gif-sur-Yvette, France}
\address {$^2$ Laboratory for Muon Spin spectroscopy, Paul Scherrer Institute, 5232 Villigen-PSI, Switzerland}
\address {$^3$Babes-Bolyai
University, Faculty of Physics, 400084 Cluj-Napoca, Romania}
\address {$^4$Service de Physique de l'Etat Condens\'e,
CEA-CNRS, CE-Saclay,  91191 Gif-Sur-Yvette, France}
\address {$^5$Service de
Physique Statistique, Magn\'etisme et Supraconductivit\'e,
CEA-Grenoble, 38054 Grenoble, France}
\address{$^6$Institut La\"ue Langevin, 6 rue Jules Horowitz, BP 156X, 38042
Grenoble, France}
\address{$^7$European Synchrotron Radiation Facility, BP 220,
38043 Grenoble, France}

\begin{abstract}

R$_2$Mo$_2$O$_7$ compounds show a ferromagnetic metal-insulator
spin glass transition tuned by the radius of the rare earth ion
R$^{3+}$. We have studied Gd$_2$Mo$_2$O$_7$ located on the verge
of the transition, by neutron diffraction on a $^{160}$Gd isotopic
sample, $\mu$SR and X ray diffraction using the synchrotron
radiation. All measurements were done both at ambient and under
applied pressure. At ambient pressure, a ferromagnetic state is
observed below the Curie temperature (T$_{\rm C}$ = 70 K). The
ordered magnetic moments at 1.7 K are parallel and equal to 5.7(5)
$\mu_{\rm B}$ and 0.8(2) $\mu_{\rm B}$ for Gd and Mo,
respectively. The relaxation rate measured by $\mu$SR evidences
strong spin fluctuations below T$_{\rm C}$ and down to the lowest
temperature (6.6 K). A spin reorientation occurs in the range 20
K$<$T$<$T$_{\rm C}$. The ferromagnetic state is strongly unstable
under pressure. T$_{\rm C}$ sharply decreases (down to 38 K at 1.3
GPa), and
 Bragg peaks start to coexist
with mesoscopic ferromagnetic correlations. The ordered moments
decrease under pressure. At 2.7 GPa long range magnetic order
completely breaks down. In this spin glass state, Gd-Gd spin
correlations remain ferromagnetic with a correlation length
limited to the fourth neighbor, and Gd-Mo spin correlations turn
to antiferromagnetic. The unique combination of three microscopic
probes under pressure provides a detailed description of the
magnetic transition, crucial for further theories.

\end{abstract}

 \maketitle

\section{Introduction}

Pyrochlores compounds R$_2$T$_2$O$_7$, where R$^{3+}$ and
T$^{4+}$are rare earth and transition or sp metal ions
respectively, show geometrical frustration of the first neighbor
interactions. This occurs not only for antiferromagnetic
interactions between first neighbor Heisenberg moments, but also
for ferromagnetic interactions if the neighboring moments are
constrained to lie along their local Ising anisotropy axes. This
peculiarity of the pyrochlore lattice leads to exotic types of
short range magnetic orders, such as spin liquids, spin ices or
chemically ordered spin glasses, which are intensively studied
\cite{Bramwell012,Greedan01}.

Among the pyrochlores, rare earth molybdates R$_2$Mo$_2$O$_7$ have
attracted special attention since the discovery of a crossover
transition from an insulating spin glass state to a metallic
ferromagnetic state, which can be tuned by the rare earth ionic
radius\cite{Katsufuji00, Morimoto01}. The variation of the
transition temperature which shows a universal curve for all
studied compounds with mixed rare earth ions, suggests that the
dominant mechanism comes from a change in the sign of the Mo-Mo
interactions. This change is connected with a change in the band
structure, due to the specific energy of Mo  t$_{2g}$ orbitals,
situated nearby the Fermi level and well separated from the other
bands \cite{Kang02, Solovyev03}. The insulating state observed at
low rare earth  ionic radius is attributed to the opening of a
Mott-Hubbard gap at the Fermi level, due to strong electronic
intrasite interactions\cite{Solovyev03}. With increasing the rare
earth ionic radius, Mo-Mo first neighbor interactions seem to
evolve from antiferromagnetic type, dominated by a superexchange
mechanism and frustrated by the geometry, to a ferromagnetic type,
due to a double exchange mechanism.

In the ferromagnetic region, the ferromagnetic alignment of the
neighboring Mo moments favors the electronic kinetic energy and a
metallic conductivity, with a mechanism similar to that observed
in the manganites. The rare earth crystal field anisotropy also
plays a role, since it yields a possible source of frustration in
the ferromagnetic region. Such frustration seems to have important
consequences on the conductivity properties. Especially, the
compound Nd$_2$Mo$_2$O$_7$ shows a giant abnormal Hall effect at
low temperature\cite{Taguchi01,Taguchi03} which cannot be
explained by the spin-orbit coupling. It is generally admitted
that this effect is induced by the spin ice frustration of the  4f
Nd$^{3+}$ spins, transferred to itinerant Mo electrons through f-d
interaction.

Gd$_2$Mo$_2$O$_7$ is especially interesting since the Gd ionic
radius is situated just above the threshold for the metal
insulator transition. This compound was thoroughly investigated by
many techniques. Magnetic measurements\cite{Raju92,Hodges03}
suggest a ferromagnetic-like transition at T$_{\rm C}$, with
magnetic irreversibilities occuring below. The T$_{\rm C}$ values
reported by various groups range between 40 - 70 K.
M\"ossbauer measurements\cite{Hodges03} show a hyperfine field
below the transition, and the hyperfine populations are found to
be out of thermal equilibrium at 27 mK, indicating that Gd and Mo
spin fluctuations are present in the magnetic phase and persist
down very low temperature. Heat capacity
measurements\cite{Schnelle04} show two step-like anomalies,
suggesting two magnetic transitions, one at T$_{\rm C}$ (70 K) and
the other well below (11 K). The transport properties strongly
depend on sample preparation. First measurements on powdered
samples \cite{Greedan87} show a metallic conductivity, with
anomalies at T$_{\rm C}$ and in the 10 K - 20 K temperature range.
However, more recent data on high purity single
crystals\cite{Kezsmarki04} show an insulating ground state, which
is very sensitive to impurity doping. In contrast with
Nd$_2$Mo$_2$O$_7$, the Hall conductivity \cite{Taguchi03} shows
the behavior of a conventional ferromagnet, as expected for
Heisenberg like Gd$^{3+}$ moments (4f$^ 7$) without orbital
angular momentum. An investigation by Infra Red (IR)
spectroscopy\cite{Kezsmarki04} shows the presence of a very small
Mott-Hubbard gap of 20 meV, about 10 times smaller than in the
neighboring insulating compounds (170 and 250 meV for Dy and Ho,
respectively). The concomitant changes in the optical, magnetic
and transport properties suggest a quantum phase transition around
Gd$_2$Mo$_2$O$_7$, opening the possibility to tune the transition
by pressure and/or magnetic field. Magnetic measurements under
pressure\cite{Park03} and recent resistivity
measurements\cite{Hanasaki06} demonstrated this possibility at a
macroscopic level.

Up to now, there was no characterization of the type of magnetic
order and magnetic fluctuations in Gd$_2$Mo$_2$O$_7$. This is
especially due to the huge absorption of natural Gd, which makes
neutron scattering experiments extremely difficult. We present
here an investigation of the magnetic order and magnetic
fluctuations in Gd$_2$Mo$_2$O$_7$, by combining neutron
diffraction on a $^{160}$Gd isotopic sample with $\mu$SR
experiments. We show that a ferromagnetic collinear order is
indeed stabilized well below T$_{\rm C}$, but that it coexists
with low temperature fluctuations. At intermediate temperatures
(20 K$<$T$<$T$_{\rm C}$) a reorientation of the magnetic moments
occurs.

\begin{figure} [h]
\includegraphics* [width=\columnwidth] {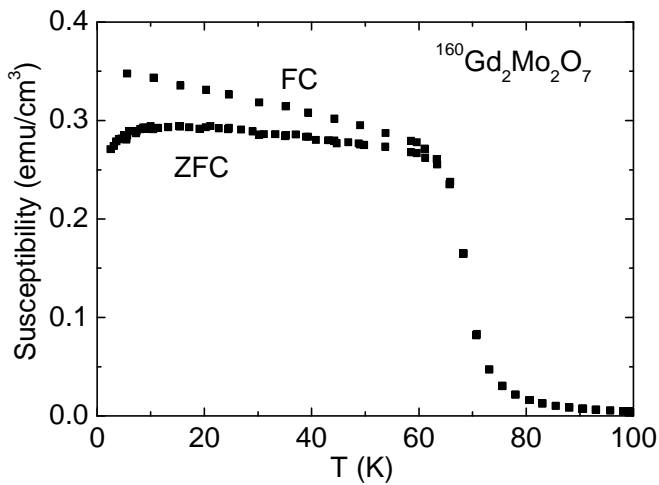}
\caption{Magnetic susceptibility measured in a static field of 50
G, in the zero field cooling (ZFC) and field cooling (FC)
processes.} \label{fig1.eps}
\end{figure}

We also studied the sensitivity of the magnetic state in
Gd$_2$Mo$_2$O$_7$ to applied pressure by neutron diffraction and
$\mu$SR. The pressure induced changes of the crystal structure and
lattice constant were checked by X ray diffraction using the
synchrotron radiation. We observe a pressure induced magnetic
transition, which can be fully characterized using this unique
combination of three microscopic probes. We show that the
ferromagnetic state is highly unstable under pressure : a rather
small pressure of 1.3 GPa yields a decrease of the Curie
temperature by a factor 2. At 2.7 GPa, long range ferromagnetic
order has fully disappeared. Our results demonstrate that the
changes in magnetism induced by pressure are equivalent to those
induced by chemical pressure (R substitution). They confirm at a
microscopic level the conclusions inferred from the
magnetization\cite{Park03}. Therefore, keeping the same sample, a
quantitative analysis of the neutron and muon data allows us to
follow the evolution of the spin correlations and fluctuations
throughout the transition.

The paper is organized as follows. In section II, we describe the
sample characterization, bulk magnetic properties and crystal
structure. In section III, we describe the magnetic state at
ambient pressure, as studied from neutron diffraction and $\mu$SR
versus temperature. In section IV, we show the evolution of the
magnetic state under pressure from X ray, neutron and $\mu$SR
experiments. In section V, we discuss the results in comparison
with other experimental data and current theories.

\begin{figure} [h]
\includegraphics* [width=\columnwidth] {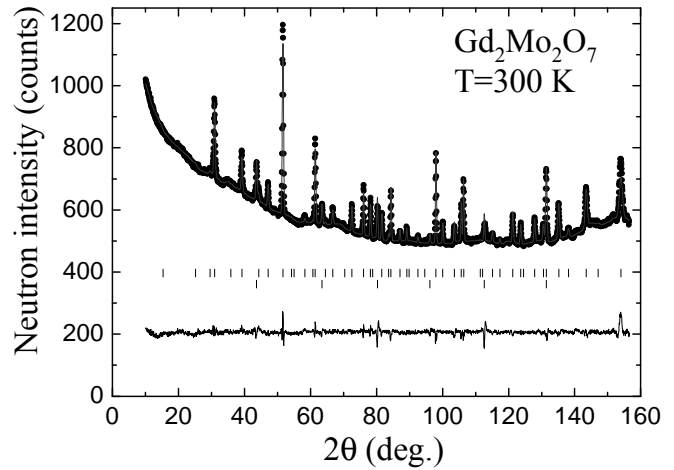}
\caption{Neutron diffraction pattern of Gd$_2$Mo$_2$O$_7$ measured
at 300 K (paramagnetic phase) on the D2B powder diffractometer,
with an incident neutron wavelength of 1.594 \AA. Upper and lower
tick marks show the Bragg peak positions of the sample and
vanadium sample holder, respectively. Solid lines show the
calculated pattern ({\itshape R$_B$}= 8.9\%) and the difference
spectrum.} \label{fig2.eps}
\end{figure}

\section{Crystal structure and bulk magnetic properties}

Powdered Gd$_2$Mo$_2$O$_7$ samples were synthesized following the
procedure given in ref \onlinecite{Katsufuji00}. We used
Gd$_2$O$_3$ and MoO$_2$ oxides as starting materials. The
synthesis was made in Argon atmosphere without any excess of Mo,
but with Ti/Zr chips to absorb oxygen traces. Samples were heated
up to 1370 $^{\circ}$C during 6 hours. Two annealings at least
were necessary to obtain the sample in pure form. A sample of 1 g
with natural Gd was prepared for synchrotron and $\mu$SR
experiments and a sample of 0.5 g with isotopically enriched
$^{160}$Gd was used for the neutron experiments. Both samples were
characterized by X ray diffraction
and magnetization. They are single phase 
and show the same magnetic transition.

The DC magnetization (Fig. 1) was recorded in a SQUID magnetometer
with a static field of 50 G. It shows a rather broad Curie
transition, as already reported in the literature. The T$_{\rm C}$
value of 70 K reported below corresponds to the inflexion point of
the susceptibility curve, whereas the onset of the transition is
situated around 80 K. Irreversibilities of the susceptibility
depending on the cooling process, -zero field cooling (ZFC) or
field cooling (FC)-, are observed below the transition. The FC
susceptibility linearly increases with decreasing temperature,
whereas the ZFC susceptibility flattens and starts to drop below
about 14 K. This drop suggests a blocking of domains walls
mobility, possibly related to the spin reorientation discussed
below.

The crystal structure of Gd$_2$Mo$_2$O$_7$ was investigated at 300
K by combining powder X-ray and neutron diffraction, the neutron
pattern being measured in the high resolution-high flux
diffractometer D2B of the Institute La\"ue Langevin with an
incident wavelength of 1.594 \AA\ in the high intensity version.
In order to decrease the residual absorption,
the sample was placed in a hollow vanadium cylindrical container. 
In the refinement, we made a specific absorption correction, taking
into account the container geometry \cite{Schmitt},
the neutron wavelength, and the isotopic composition of the sample.
The linear absorption coefficients were estimated to $\mu$ = 6.3 and
4.1 cm$^{-1}$ for D2B and D20, respectively.

The Rietveld refinement of the D2B diffraction pattern (Fig. 2)
was made  using the crystallographic programs of the Fullprof
suite\cite{Carvajal93}. We took the diffraction pattern from the
vanadium container into account as a second phase. The refinement
({\itshape R$_B$}=8.9 \%, {\itshape R$_F$}=5.9 \% ) confirmed the
structural model for a stoechiometric pyrochlore with space group
{\itshape Fd$\overline{3}$m}, yielding a lattice constant
{\itshape a}= 10.3481(2) \AA\ and an oxygen position parameter
{\itshape u}= 0.3342(2) at room temperature.
These results agree with previous determinations from synchrotron radiation X ray powder diffraction\cite{Morimoto01}. 

\begin{figure} [h]
\includegraphics* [width=\columnwidth] {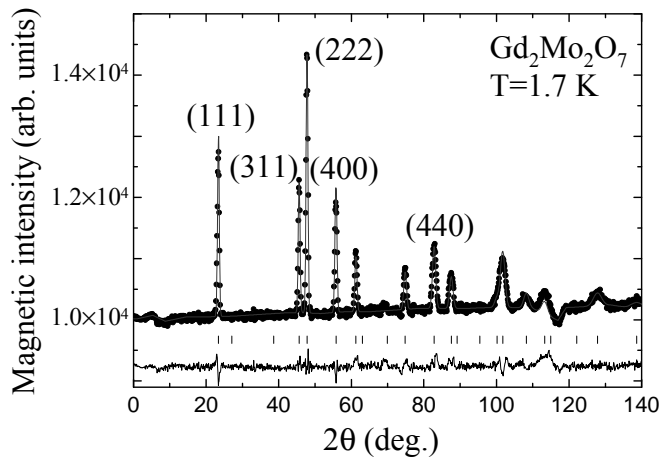}
\caption{Magnetic neutron diffraction pattern of Gd$_2$Mo$_2$O$_7$
measured at 1.7 K on the D20 powder diffractometer, with an
incident neutron wavelength of 2.419 \AA. A pattern measured at 90
K ( above T$_{\rm C}$) was subtracted. Tick marks show the Bragg
peak positions. Solid lines show the pattern calculated with the
collinear ferromagnetic model ({\itshape R$_B$}= 11.3 \%) and the
difference spectrum.} \label{fig3.eps}
\end{figure}

\section{Magnetic state at ambient pressure}
\subsection{Magnetic neutron diffraction}

Neutron diffraction patterns were recorded between 1.7 K and 90 K
on the high flux diffractometer D20 of the Institute La\"ue
Langevin with an incident wavelength 2.419 \AA, with the same
sample and sample holder as for D2B. Absorption corrections were
made as above. Magnetic diffraction patterns were obtained by
subtracting a spectrum at 90 K just above the transition. The
magnetic pattern at 1.7 K is shown in Fig. 3. The magnetic peaks
have $\it{hkl}$ indices of the face centered cubic lattice. The
(200) and (220) peaks, where there is no chemical contribution
from Gd and Mo ions in the pyrochlore structure (due to extinction
from the {\itshape Fd$\overline{3}$m} space group and special
Gd-Mo positions, respectively) are also absent in the magnetic
pattern. We refined the magnetic patterns using the Fullprof
suite\cite{Carvajal93}. At 1.7 K, we obtain a good refinement
assuming a collinear ferromagnetic structure (magnetic factor
{\itshape R$_B$}= 11.3 \%, Fig. 3). The Mo and Gd moments orient
along the same direction, with a ferromagnetic coupling. According
to the $^{155}$Gd M\"ossbauer measurements\cite{Hodges03}, the
angle between the Gd moment and the $<$111$>$ axis is close to
54$^\circ$. This strongly suggests that the Gd and Mo moments lie
along a $<$100$>$ axis.

With increasing temperature the intensities of the magnetic Bragg
peaks start to decrease, but the (111) peak decreases much more
rapidly than the high angle peaks (Fig. 4). It disappears around
50 K whereas the other peaks persist up to 70 K. Concomitantly,
the magnetic intensities calculated within the collinear
ferromagnetic model start to disagree with the experimental data
(inset Fig. 4, {\itshape R$_B$}= 39 $\%$ at 40 K).

The refinements exclude a zero ordered moment on either Gd or Mo
sublattice.
A global change of orientation of both Gd and Mo sublattices is
also excluded since it would not change the magnetic diffraction
pattern of a powdered cubic structure. The observed anomaly is
therefore attributed to a decoupling of Gd$^{3+}$ and Mo$^{4+}$
moments above 20 K, yielding non collinear moments in the Gd (and
possibly Mo) sublattice. Unfortunately, a precise determination of
this non-collinearity is hampered by the high symmetry of the
crystal structure, the presence of eight independent moments in
the unit cell, and the dominant ferromagnetic character.

To search for non collinear structures, we first performed a
symmetry analysis\cite{Bertaud68,Izyumov91} using the program
BasIreps\cite{BASIREPS}, searching for all {\bf k}=0 structures
corresponding to irreducible representation of the {\itshape
I4$_1$$/$amd} space group, the highest symmetry group allowing
ferromagnetism. We also checked non collinear structures predicted
by theory, assuming either anisotropic or dipolar
interactions\cite{Champion02,Palmer00}. Finally, we explored small
deviations from the ferromagnetic collinear case using the
simulated annealing process available in
Fullprof\cite{Carvajal93}. Whatever the procedure, we could not
improve the quality of the refinement significantly with respect
to the collinear ferromagnetic model. We conclude that a non
collinearity of the Gd/Mo ordered moments indeed occurs above 20
K, although we cannot determine its nature presently. We also
notice a slight change of the peak linewidth, showing that the
magnetic correlation length increases with decreasing T in this
temperature range (inset Fig. 5).

\begin{figure} [h]
\includegraphics* [width=\columnwidth] {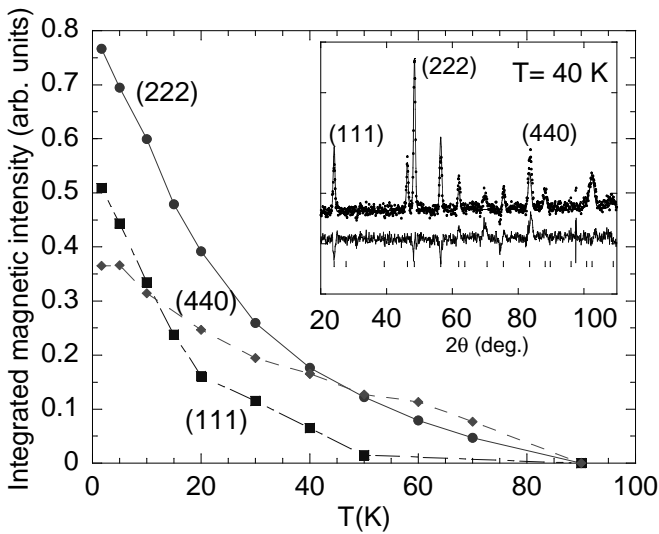}
\caption{Integrated intensity of several magnetic Bragg peaks in
Gd$_2$Mo$_2$O$_7$, measured versus temperature on D20. The
intensities are scaled to the intensity of the (222) nuclear peak
at 90 K. In inset the magnetic diffraction pattern at 40 K. }
\label{fig4.eps}
\end{figure}

The temperature dependence of the ordered magnetic moments deduced
from refinements in the collinear ferromagnetic model is plotted
in Fig. 5. At 1.7 K, the moments values deduced from the
refinement are 5.70(8) $\mu_{\rm B}$ and 0.82(5) $\mu_{\rm B}$ for
Gd and Mo, respectively. Considering the uncertainty on the scale
factor between nuclear and magnetic scattering due to the
absorption correction, we estimate the ordered magnetic moments in
absolute scale with a larger error bar, namely 5.7(5) $\mu_{\rm
B}$ and 0.8(2) $\mu_{\rm B}$ for Gd$^{3+}$ and Mo$^{4+}$ ions,
respectively. The ordered Gd$^{3+}$ moment is reduced with respect
to the free ion value of 7 $\mu_{\rm B}$. The ordered Mo$^{4+}$
moment is also much smaller than the value of 2 $\mu_{\rm B}$
expected assuming a ionic description (4d$^2$ t$_{2g}$ state with
S=1 and g=2). The calculated ordered moment per formula unit of
13(1) $\mu_B$ is significantly smaller than the saturated value of
16.8 $\mu_B$ obtained at 2 K in a high field of 14 T
\cite{Hanasaki06}. Since we do not expect any strong reduction of
the free ion values from crystal field effects (the Gd$^{3+}$
moment has no orbital contribution), the reduced moment values
should be associated with the low temperature fluctuations. This
is also supported by the $\mu$SR experiments, as discussed in the
next section.

\begin{figure} [h]
\includegraphics* [width=\columnwidth] {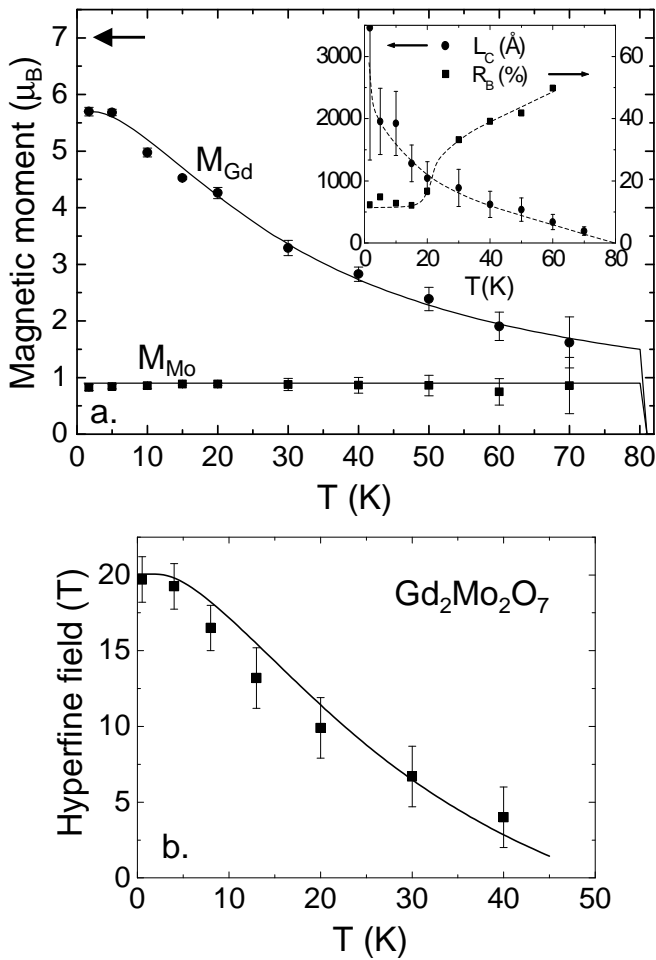}
\caption{a:Temperature dependence of the ordered moments M$_{Gd}$
(circles) and M$_{Mo}$ (squares) in Gd$_2$Mo$_2$O$_7$ deduced from
refinements of the neutron data in the collinear ferromagnetic
model. The arrow shows the free ion value of 7 $\mu_{\rm B}$. The
solid line for Gd correspond to a fit with a Brillouin function,
as described in the text. In inset, the magnetic correlation
length L$_c$ deduced from the peak widths (dots) and the magnetic
factor {\itshape R$_B$} (squares) versus temperature. Dashed lines
are guides to the eye. b:Thermal variation of the $^{155}$Gd
hyperfine field in Gd$_2$Mo$_2$O$_7$, from Ref. \cite{Hodges03}.
The line is a fit to the sum of a contact hyperfine field,
proportional to the Gd moment, and of a transferred field,
proportional to the Mo moment, and thus temperature independent.}
\label{fig5.eps}
\end{figure}

In Fig. 5a, we fitted M$_{Gd}$(T) by a Brillouin function
B$_{7/2}$, assuming that the Gd ions are submitted to a molecular
field coming from the Mo ions. As the Mo spontaneous moment is
practically temperature independent up to T$_{\rm C}$, we assumed
that the molecular field H$_{ex}$ is constant up to 80 K and
vanishes above. The fit in Fig. 5 is obtained  with H$_{ex}$ =
10.8
T.

 The thermal variation of the Gd moment
has also been inferred from the $^{155}$Gd Mössbauer data in
Gd$_2$Mo$_2$O$_7$ in Ref. \onlinecite{Hodges03}, assuming that the
Gd moment is proportional to the hyperfine field. Comparison with
the present neutron data shows a clear disagreement: the
M\"ossbauer derived values fall much more rapidly than the neutron
values as temperature increases. So we propose a new
interpretation of the M\"ossbauer data: the measured hyperfine
field H$_{\rm hf}$ is assumed to be the sum of a contact hyperfine
field, proportional to the Gd moment, and of a transferred
hyperfine field H$_{\rm tr}$ coming from the polarization of the
conduction band by the Mo moments
\cite{Tomala77,Dormann91}:$\hspace{\fill}$

\begin{equation}
        H_{\rm hf}(T) = \vert A M_{Gd}(T) + H_{\rm tr}
        \vert
\end{equation}

 The sign of H$_{\rm hf}$ cannot be obtained by
the M\"ossbauer data. The transferred field is proportional to the
Mo moment, and it is thus temperature independent up to T$_{\rm
C}$. In Fig. 5b, we have represented the hyperfine field data from
Ref. \onlinecite{Hodges03} and the fit to the above formula, where
M$_{Gd}$(T) is calculated using the Brillouin function as above.
The fit yields a contact hyperfine constant A=5.8 T/$\mu_{\rm B}$
and a transferred field  H$_{\rm tr}$= - 13 T. The sign of the
latter is opposite to that of the contact term, which can occur in
intermetallic Gd compounds with 3d or 4d metals
\cite{Tomala77,Dormann91}. Above 40 K, the measured hyperfine
field is below 4 T, with a large error bar. The above law predicts
that H$_{\rm hf}$(T) shows a minimum at 50 K (not shown), which
could not be observed due to the small hyperfine field values.
According to this picture, Gd$_2$Mo$_2$O$_7$ is an example where
the measured $^{155}$Gd hyperfine field is not proportional to the
spontaneous moment.

\subsection{$\mu$SR measurements}

$\mu$SR measurements at ambient pressure were performed at the Swiss
Muon Source at the Paul Scherrer Institute (Villigen, Switzerland)
on the GPS instrument in the temperature range 6.6 K - 300 K.
Selected $\mu$SR spectra are shown in Fig. 6. Above T$_{\rm C}$, the
relaxation function P$_{Z}$(t) shows an exponential like decay:
P$_Z$ (t)= exp(-$\lambda$$_Z$t). Below T$_{\rm C}$, P$_{Z}$(t) shows
a rapidly damped oscillation at early times which is readily
attributed to the presence of magnetic order. An exponential decay
is also observed at long time scales in the ordered phase, showing
that spin fluctuations persist down to the lowest temperature.

\begin{figure} [h]
\includegraphics* [width=\columnwidth] {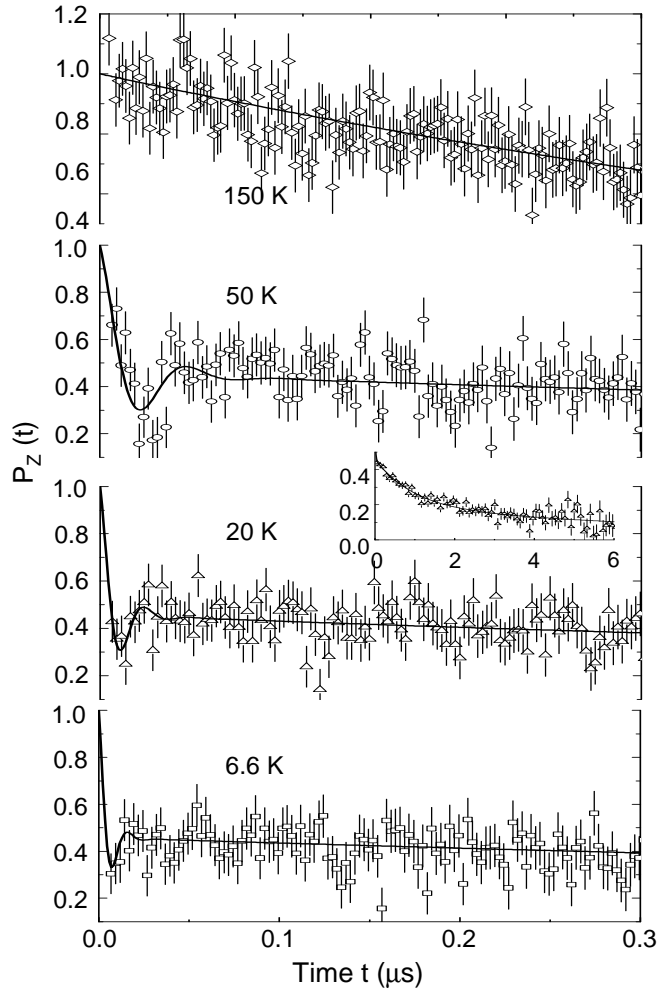}
\caption{Muon spin depolarization function P$_Z$(t) in
Gd$_2$Mo$_2$O$_7$ measured on the GPS beam line at several
temperatures (ambient pressure). Lines are fits as described in
text. The inset shows the depolarization function in an extended
time range for T=20 K.} \label{fig6.eps}
\end{figure}

Below T$_C$, the relaxation function P$_{Z}$(t) was fitted by the
equation: 

\begin{equation}
P_Z(t)= [exp(-\lambda_Zt) + 2 exp(-\lambda_Tt)
cos(\gamma_{\mu}<B_{loc}>t)]/3.
\end{equation}


This equation is expected to hold in the magnetically ordered
phase of a powder sample\cite{Dalmas04}. The first term
corresponds to the depolarization by spin fluctuations with the
longitudinal relaxation rate $\lambda$$_{Z}$, whereas the second
term reflects the precession of the muon spin in the average local
field $<$B$_{loc}$$>$ at the muon site. $\gamma$$_\mu$ is the muon
gyromagnetic ratio. The transverse relaxation rate $\lambda$$_{T}$
can have both static and dynamical character. In the high
temperature limit, when the rate of fluctuations of Gd$^{3+}$ and
Mo$^{4+}$ moments is much larger than the coupling between the
muon spin and the electronic spins, one gets $<$B$_{loc}$$>$=0,
and $\lambda$$_{T}$=$\lambda$$_{Z}$. Then the
 above equation reduces to a simple exponential P$_Z$(t)=
exp(-$\lambda$$_Z$t). This simple function was used to fit the data
above T$_C$.

\begin{figure} [h]
\includegraphics* [width=\columnwidth] {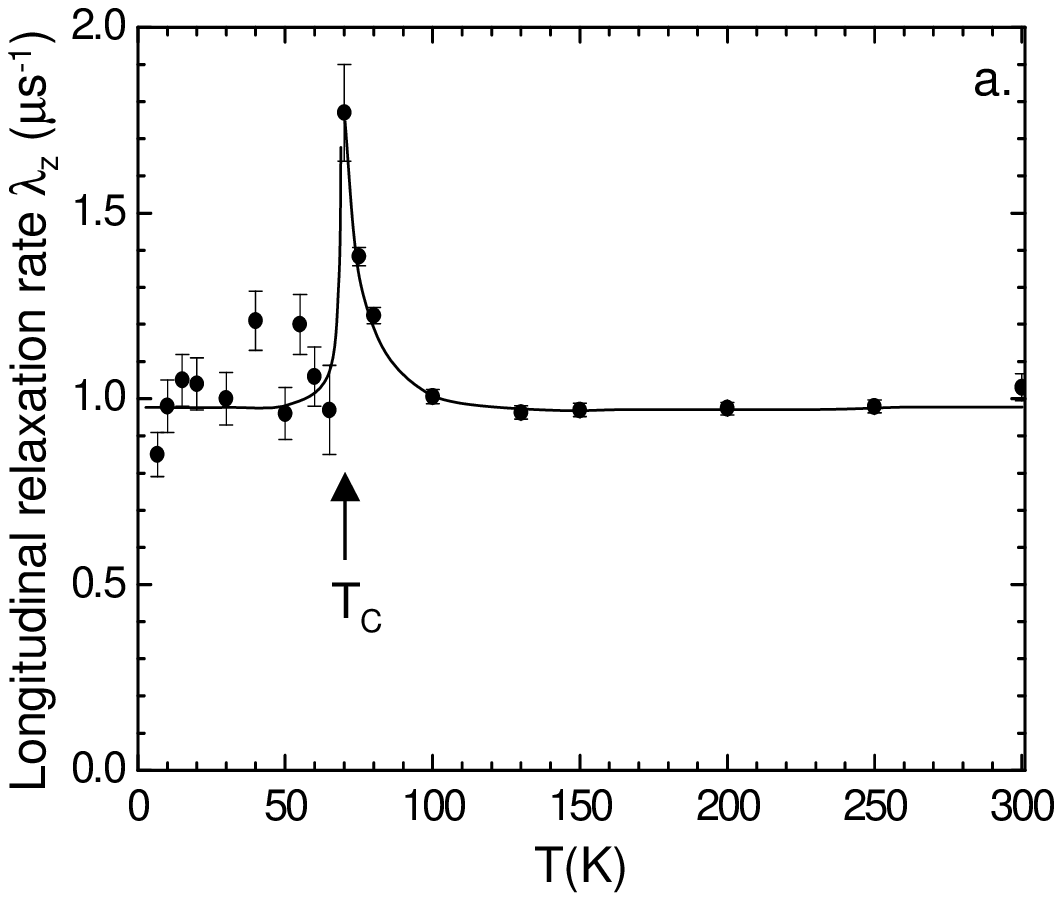}
\includegraphics* [width=\columnwidth] {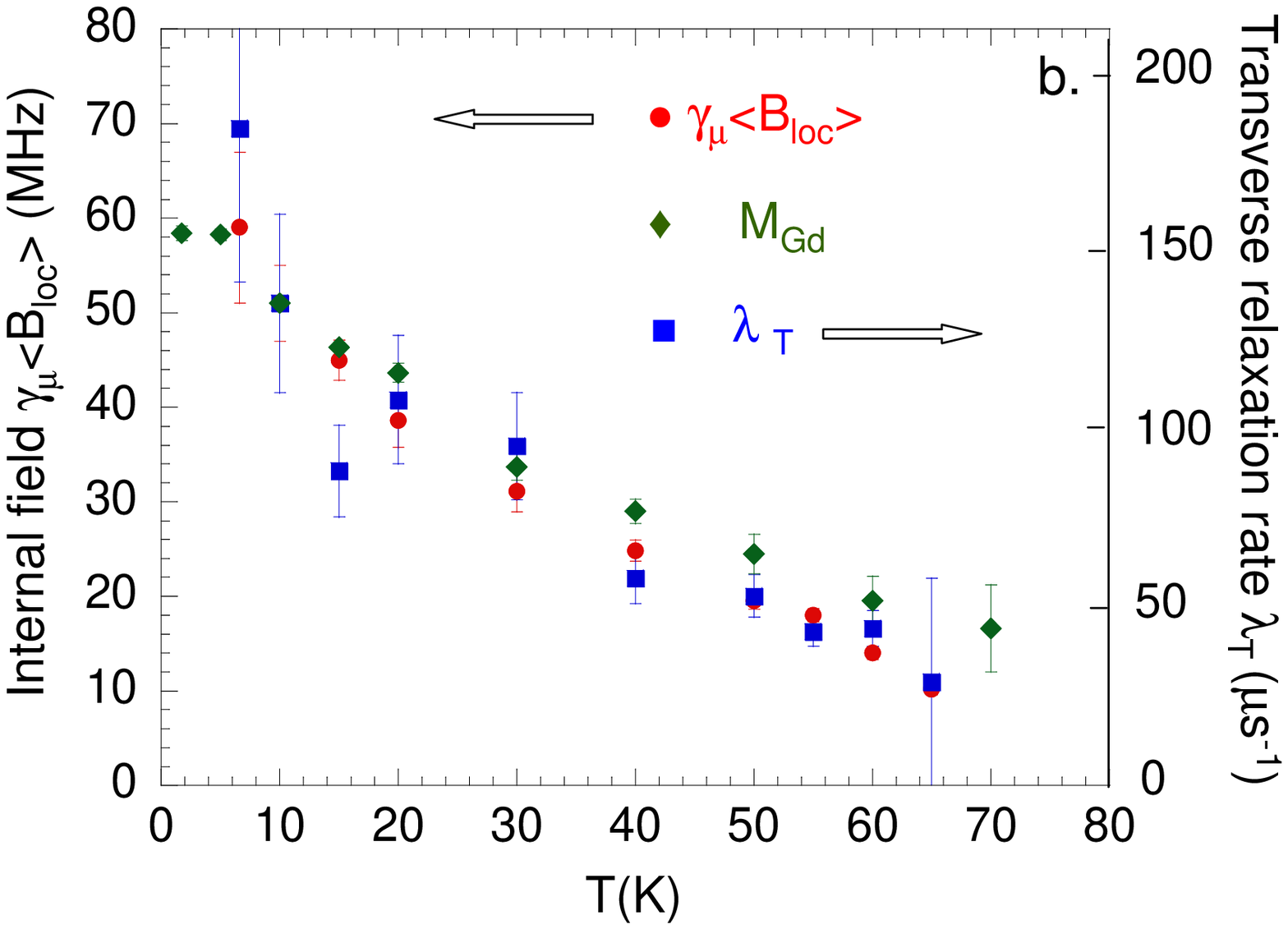}
\caption{From $\mu$SR measurements in Gd$_2$Mo$_2$O$_7$. a:
temperature dependence of the longitudinal relaxation rate
$\lambda$$_{Z}$; b: temperature dependence of
$\gamma$$_{\mu}$$<$B$_{loc}$$>$ (with $\gamma$$_\mu$ the muon
gyromagnetic ratio), $\lambda$$_{T}$ and M$_{Gd}$ (scaled). }
\label{fig7.eps}
\end{figure}

In a standard ferromagnet, the longitudinal relaxation rate
$\lambda$$_{Z}$ shows a cusp at the critical temperature, then
decreases with decreasing temperature below T$_{\rm C}$ and tends
to zero as $T\to0$. In contrast, in Gd$_2$Mo$_2$O$_7$,
$\lambda$$_{Z}$ keeps a high value down to the lowest temperature
(6.6 K), and remains T-independent in the whole ordered phase
(Fig. 7a) almost up to T$_{\rm C}$. This abnormal behavior
reflects the presence of strong spin fluctuations coexisting with
the magnetic order, which seem to persist as $T\to0$ in contrast
with standard spin waves. Such peculiar fluctuations have already
been observed by $\mu$SR in several geometrically frustrated
magnets with a pyrochlore lattice. They were seen in spin liquids
\cite{Gardner99,Bonville04} like Tb$_2$Ti$_2$O$_7$ and
Yb$_2$Ti$_2$O$_7$, as well as ordered compounds
\cite{Bonville04,Bert06,Dalmas06} like Gd$_2$Sn$_2$O$_7$ and
Tb$_2$Sn$_2$O$_7$. In all cases, the longitudinal relaxation rate
saturates in the lowest T range (to values around 1$\mu$s$^{-1}$),
as here for Gd$_2$Mo$_2$O$_7$.

The transverse relaxation rate $\lambda$$_{T}$ is up to two orders
of magnitude larger than $\lambda$$_{Z}$(T). It smoothly increases
below T$_C$ and scales with the average local field
$<$B$_{loc}$$>$ as temperature varies. This leads us to assign
$\lambda$$_{T}$ mainly to the width of the distribution of local
fields. We notice that both quantities also scale with the ordered
moment M$_{Gd}$(T) determined by neutron diffraction (Fig. 7b). It
suggests that the local field seen by the muon mostly comes from
the Gd$^{3+}$ ions with much larger moments, although more
localized, than the Mo$^{4+}$ ones. Our current neutron and
$\mu$SR data in the (Tb,La)$_2$Mo$_2$O$_7$ series \cite{Apetrei06}
as well as previous $\mu$SR data in Tb$_2$Mo$_2$O$_7$ and
Y$_2$Mo$_2$O$_7$ spin glasses \cite{Dunsiger96} support this
interpretation, showing that the static internal field is about 10
times higher in Tb$_2$Mo$_2$O$_7$ with both Tb$^{3+}$ and
Mo$^{4+}$ moments than in Y$_2$Mo$_2$O$_7$ where only Mo$^{4+}$
moments are involved.

\section{The pressure induced state}
\subsection{Crystal structure under pressure : X ray diffraction}

X-ray diffraction under pressure using the synchrotron radiation
was performed at room temperature on the ID27 beam line of ESRF,
in the pressure range 0-10 GPa, with an incident wavelength of
0.3738 $\AA$. We used a diamond anvil cell and an ethanol-methanol
mixture as pressure transmitting medium. The crystal structure
remains cubic with {\itshape Fd$\overline{3}$m} space group in the
whole measured pressure range.

The evolution of the lattice constant {\itshape a} versus pressure
is shown in Fig. 8. From ref. \onlinecite{Katsufuji00}, one
estimates the critical lattice constant for the ferromagnetic-spin
glass transition to {\itshape a$_c$} =10.327(5) \AA\ (critical
ionic radius {\itshape r$_c$}=1.047 \AA). Besides
Gd$_2$Mo$_2$O$_7$, we also studied two samples under pressure in
comparison, namely Tb$_2$Mo$_2$O$_7$, an insulating spin glass
with {\itshape a$_0$}=10.312 \AA, smaller than {\itshape a$_c$},
and (Tb$_{0.8}$La$_{0.2}$)$_2$Mo$_2$O$_7$ with {\itshape
a$_0$}=10.378 \AA\ greater than {\itshape a$_c$}. In the studied
pressure range, we found the same variation of the relative volume
V/V$_0$(where V$_0$ is the unit cell volume at ambient pressure)
for all compounds.
The equation of state was fitted to the Murnaghan equation V/V$_0$
= (P*B$_1$/B$_0$ +1)$^{-1/B_{1}}$, yielding a bulk modulus
B$_0$=151.8(9) GPa . The
pressure derivative of the bulk modulus 
was fixed to a reasonable value (B$_1$= 4.5).

Knowing the evolution of the lattice constant under pressure
enables us to compare the pressure induced state in
Gd$_2$Mo$_2$O$_7$ with the ambient pressure state in
R$_2$Mo$_2$O$_7$ compounds with smaller lattice constant. As
discussed in the last section, the different experimental
determinations of the critical pressure for Gd$_2$Mo$_2$O$_7$ lie
in the pressure range 0.6-2.4 GPa.

\begin{figure} [h]
\includegraphics* [width=\columnwidth] {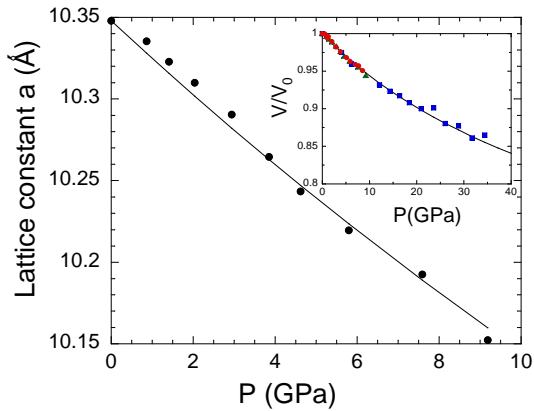}
\caption{Variation of the lattice constant {\itshape a} versus
pressure at ambient temperature. In the 
inset the relative volume V/V$_0$ is plotted versus pressure up to
40 GPa for three samples: Gd$_2$Mo$_2$O$_7$
(circles),Tb$_2$Mo$_2$O$_7$ with {\itshape a$_0$} $<$ {\itshape
a$_c$} (squares) and (Tb$_{0.8}$La$_{0.2}$)$_2$Mo$_2$O$_7$ with
{\itshape a$_0$} $>$ {\itshape a$_c$} (triangles). Solid lines are
fits to the Murnaghan equation.} \label{fig8.eps}

\end{figure}

We notice that the quality of the experimental data was not
sufficient to refine the oxygen parameter due to texture effects
and/or non isotropic powder averaging. So we cannot determine the
pressure dependence of the Mo-O-Mo bond angle.

\begin{figure} [h]
\includegraphics* [width=\columnwidth] {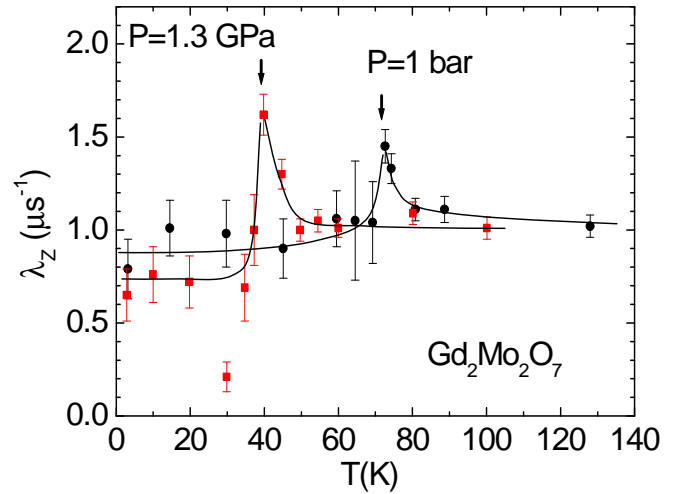}
\caption{Temperature dependence of the relaxation rate
$\lambda$$_Z$, with the Gd$_2$Mo$_2$O$_7$ sample in the pressure
cell. Data were measured on the GPD instrument: ambient pressure
(circles) and P=1.3 GPa (squares). Solid lines are guides to the
eye.} \label{fig9.eps}
\end{figure}

\subsection{Change of the transition temperature : $\mu$SR}

Muon spin rotation measurements under pressure were performed on
the GPD instrument of the Paul Scherrer Institute, using high
energy incident muons to penetrate the pressure cell. The sample
was mounted in a piston cylinder cell inserted in a cryostat. The
signal from the pressure cell (about 70 \% of the total asymmetry)
was fitted by a Kubo-Gauss function. The sample was first measured
in the pressure cell at ambient pressure in the T range 3.2 K -130
K, then the cell was pressurized and the experiment repeated. The
pressure was determined by measuring the superconducting
transition of a lead wire inside the pressure cell, yielding a
value of 1.3 GPa. Below T$_{\rm C}$ it was difficult to extract
any information from the $\mu$SR spectra at small times (the 2/3
term), due both to the large background of the pressure-cell and
to the fast depolarization of the 2/3 term. Hence, below T$_{\rm
C}$ we fitted only the 1/3 term with an exponential depolarization
function. Fig. 9 compares the temperature dependence of the
relaxation rate at ambient and under 1.3 GPa. The sharp peak in
$\lambda$$_Z$(T) clearly moves towards lower temperatures under
pressure. It reflects a very strong decrease of the transition
temperature- from 70 K to 38 K, when the pressure increases from
ambient pressure to 1.3 GPa.

\begin{figure} [h]
\includegraphics* [width=\columnwidth] {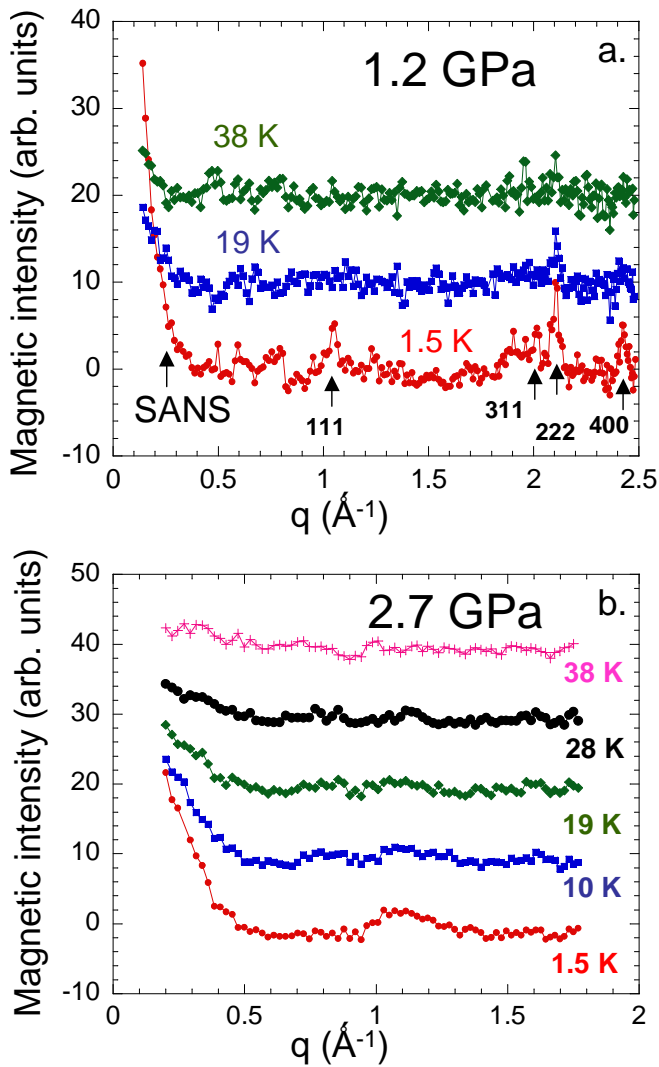}
\caption{Magnetic diffraction patterns in Gd$_2$Mo$_2$O$_7$ under
pressure versus the scattering vector {\itshape
q}=4$\pi$sin$\theta$/$\lambda$. a. P= 1.2 GPa; b. P= 2.7 GPa.
Spectra are corrected by a spectrum in the paramagnetic range (90 K)
and scaled to the nuclear intensity of the (222) Bragg peak.}
\label{fig10.eps}
\end{figure}

\subsection{Change of the spin correlations: magnetic
neutron diffraction}

Neutron diffraction measurements under pressure were performed on
the spectrometer G6-1 of the Laboratoire L\'eon Brillouin with an
incident wavelength of 4.741 \AA, in the high pressure version  
 \cite{Goncharenko04}. The sample was loaded in a sapphire anvil
cell with 40 $\%$ Al powder to decrease the sample absorption.
Measurements were performed at the pressures 0.5, 1.2, 1.9 and 2.7
GPa, between 1.5 K  and 90 K. Pressure was measured by the ruby
fluorescence technique.
The magnetic diffraction spectra were obtained by subtracting a
spectrum at 90 K (paramagnetic phase). The magnetic intensity was
scaled to the integrated intensity of the (222) Bragg peak. This
procedure allows us to compare spectra measured at different
pressures and temperatures.

Fig. 10a shows magnetic diffraction patterns at 1.2 GPa. The
spectrum at 1.5 K shows that magnetic long range (LRO) and short
range (SRO) orders coexist in the sample. Small Bragg peaks are
clearly visible, with a width limited by the experimental
resolution. An intense scattering is also observed at low
{\itshape q} values. This small angle neutron scattering (SANS),
which was absent in the ambient pressure data, shows the onset of
ferromagnetic correlations with a mesoscopic length scale. With
increasing temperature, the (111) peak disappears at about 20 K,
whereas the (222) peak vanishes at a higher temperature close to
38 K,  the transition temperature determined by $\mu$SR for this
pressure. It suggests that the spin reorientation observed in the
ordered state at ambient pressure temperature still persists under
pressure. The short range ferromagnetic correlations persist
slightly above the transition.

At 1.9 GPa, a very small ordered component remains at the Bragg
positions. At 2.7 GPa, the magnetic Bragg peaks have completely
disappeared, and only magnetic short range order is present (Fig.
10b). Short range magnetic correlations are clearly shown at 1.5
K, yielding a strongly modulated magnetic background. Here, the
SANS signal coexists with a diffuse peak centered around  1.1
\AA$^{-1}$. With increasing temperature, the diffuse peak flattens
and vanishes around 20 K, whereas the ferromagnetic correlations
persist up to 40 K.

\begin{figure} [h]
\includegraphics* [width=\columnwidth] {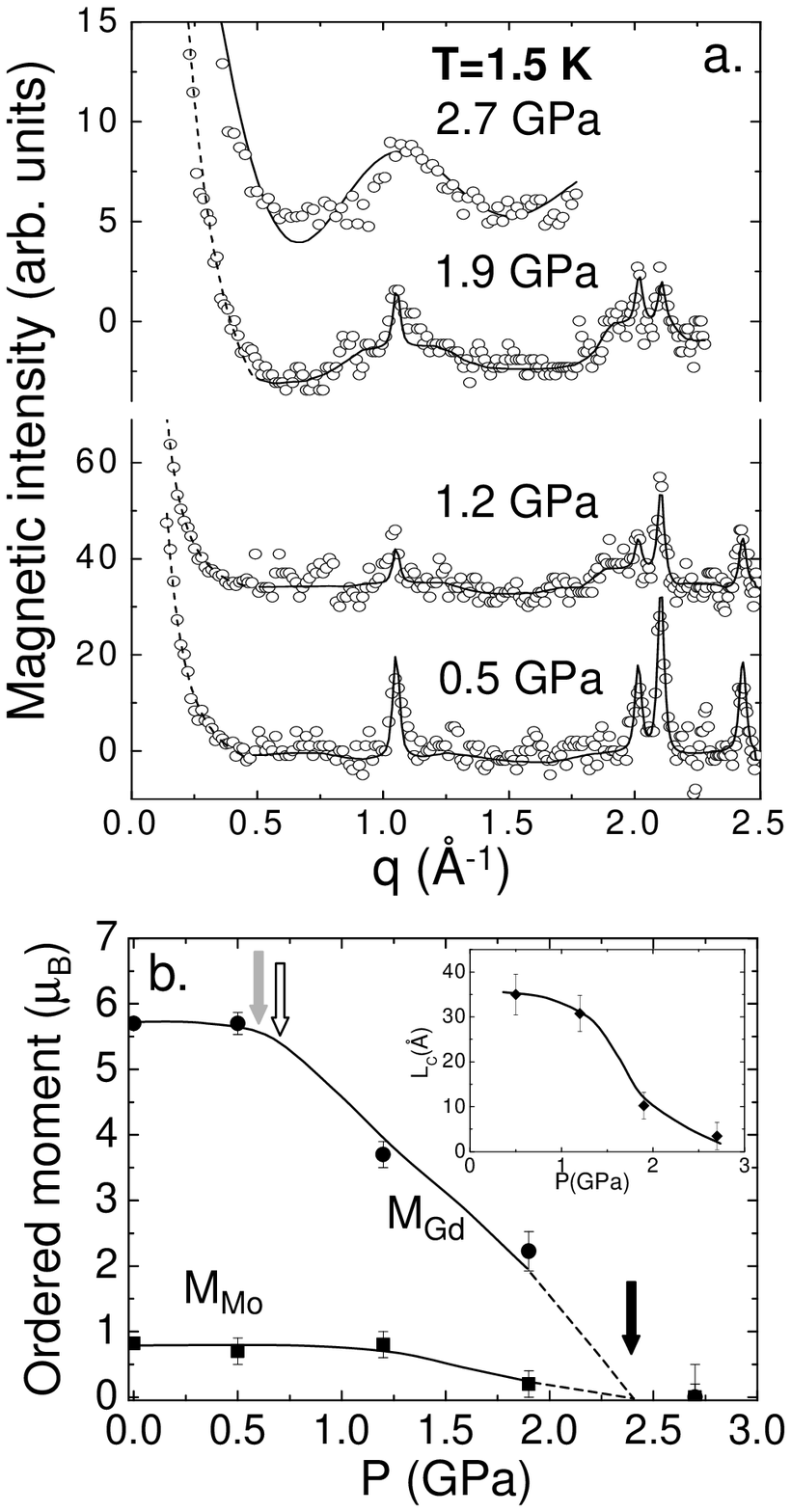}
\caption{ a. Magnetic diffraction patterns measured in
Gd$_2$Mo$_2$O$_7$ at 1.5 K for several pressures. Solid lines are
fits of the Bragg peaks (0.5, 1.2 and 1.9 GPa, {\itshape q} $>$
0.5 $\AA$$^{-1}$) and of the diffuse peaks (2.7 GPa),
respectively. The dashed lines are fits of the SANS signal (0.5,
1.2 and 1.9 GPa, {\itshape q} $<$ 0.5 $\AA$$^{-1}$). All fits are
described in text. b. Variation of the ordered moments versus
pressure. The arrows indicate the critical pressure as determined
from susceptibility\cite{Park03} (0.6 GPa, grey), chemical
pressure\cite{Katsufuji00} (0.7 GPa, white) and
resistivity\cite{Hanasaki06} (2.4 GPa, black). In inset: the
variation of the correlation length versus pressure as
obtained from the fit of the SANS.
} \label{fig11.eps}
\end{figure}

In Fig. 11a, we compare diffraction patterns at 1.5 K for several
pressures. With increasing pressure,  the decrease of the Bragg
intensity coincides with the growing of the diffuse magnetic peak,
as the relative contributions of SRO and LRO vary in opposite way.
The SANS signals flattens, showing that the ferromagnetic
correlation length decreases with increasing pressure (see inset
Fig. 11a).

To analyze the pressure data we proceeded as follows. The long
range magnetic order was analyzed within the collinear
ferromagnetic model, by refining the Bragg peaks using Fullprof.
In this procedure, the diffuse scattering and SANS signal were
treated as a background. The magnetic peaks were scaled to the
nuclear Bragg peak (222) measured in the paramagnetic range. The
calculated spectra are shown in Fig. 11a. The ordered moments are
plotted versus pressure Fig. 11b. The Gd$^{3+}$ moment strongly
decreases with increasing pressure. The Mo$^{4+}$ is expected to
decrease too, but being much smaller, its variation remains within
the experimental error.

To get information about the short range correlations yielding the
diffuse magnetic signal at 2.7 GPa, we used the model proposed by
Bertaut and Burlet\cite{Bertaut67} for spin glasses, and applied
by Greedan {\em et al.}\cite{Greedan91} to the pyrochlore system.
A fit of the diffuse magnetic scattering by the sum of radial
correlation functions was performed, giving information on
spin-spin correlation parameters $\gamma$ (Fig. 12). We considered
correlations up to the 4th neighbour shell ($\sim$ 7.3 \AA). The
correlation coefficients $\gamma$ deduced from this fit at 2.7 GPa
are plotted in inset of Fig. 12. The Gd-Gd correlations are F
($\gamma$$_{1,3,4}$$>$ 0), while the Gd-Mo are AF ($\gamma$$_2$$<$
0). The AF Mo-Mo correlations responsible for the frustration in
the SG state\cite{Gardner99} cannot be detected, their
contribution being about 50 times smaller than the Gd-Gd ones due
to the smaller Mo moment.

\begin{figure} [h]
\includegraphics* [width=\columnwidth] {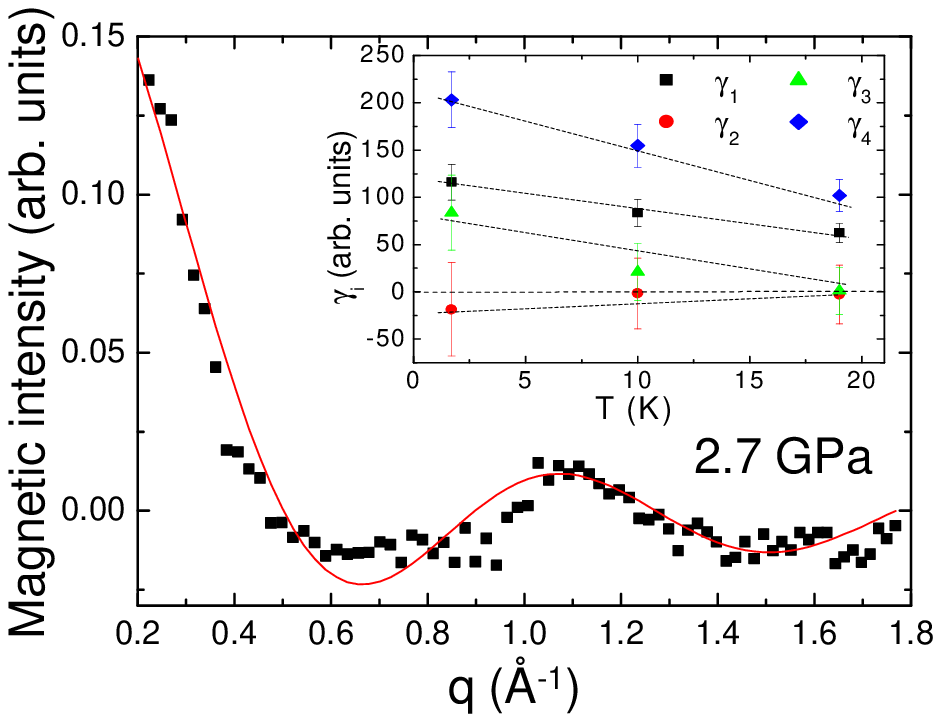}
\caption{Magnetic intensity in Gd$_2$Mo$_2$O$_7$ at 1.5 K and P=2.7
GPa. The fit is made with the short range model described in the
text. In inset the temperature dependence of the correlation
coefficients. Dashed lines are guides to the eye.} \label{fig12.eps}
\end{figure}

We outline that in the intermediate pressure range (0 $<$ P $<$
2.7 GPa) neither the ordered ferromagnetic model nor the short
range magnetic model can describe the SANS signal in a proper way.
This is because the mesoscopic length scale of the SANS is
intermediate
between 
the 4th neighbor shell (7 \AA) and the domain size associated with
the Bragg peaks (400 \AA\ or above, taking into account the
experimental resolution).
Considering the relative magnitude of the Mo and Gd local moments
and the fact that Mo-Mo and Mo-Gd correlations are AF, we
attribute the SANS signal to ferromagnetic Gd-Gd correlations. To
get an insight about their correlation length, we have fitted the
SANS with a Lorentzian function. The correlation length L$_{\rm
c}$ is about 35 \AA\ at 0.5 GPa and 1.5 K. It decreases with
increasing pressure (inset Fig. 11b) and temperature.

\section{Discussion}

In this section, we discuss the main results in comparison with
other experimental data and existing theories. At ambient
pressure, the ordered ground state in  Gd$_2$Mo$_2$O$_7$ is a
collinear ferromagnet, in contrast with Nd$_2$Mo$_2$O$_7$, where
the Nd$^{3+}$ free ion anisotropy induces a spin ice configuration
of the Nd$^{3+}$ moments\cite{Taguchi01,Yasui03}.

A collinear ground state is expected in Gd$_2$Mo$_2$O$_7$ from the
band structure\cite{Solovyev03}, and from the spin only character
of Gd$^{3+}$ ion. It could explain why there is no giant anomalous
Hall effect, in contrast with Nd$_2$Mo$_2$O$_7$. The ordered
magnetic state is however strongly abnormal, even at ambient
pressure. It coexists with strong fluctuations, which we probe by
$\mu$SR, and which persist down to 27 mK according to M\"ossbauer
data\cite{Hodges03}. These fluctuations could be responsible for
the strong reduction of the ordered moments. The Gd$^{3+}$ moment
is reduced by 20$\%$, and the Mo$^{4+}$ moment by 60 $\%$ with
respect to the free ion values. Spin fluctuations likely play a
role in the spin non collinearity above 20 K. The origin of these
fluctuations needs to be clarified. Low temperature collective
fluctuations are widely observed in geometrically frustrated
magnets and seem to be a key feature of geometrical
frustration\cite{Bramwell01}. However, in Gd$_2$Mo$_2$O$_7$ at
ambient pressure, both Gd-Gd and Mo-Mo first neighbor interactions
are ferromagnetic, and should be frustrated only by the anisotropy
of Gd or Mo ions. This anisotropy is very small, and obviously
does not play a key role to select the ground state.
Alternatively, one could speculate that the abnormal fluctuations
come from the metal-insulating instability. Taking them into
account in the band structure is a challenge for theory.

The ferromagnetic state is strongly unstable under pressure, so
that a small pressure of 0.5 GPa induces important changes in the
magnetic correlations and strongly decreases the T$_{\rm C}$
value.
Actually in Gd$_2$Mo$_2$O$_7$ our neutron data show that the
magnetic order gradually evolves from F to SG in the pressure range
of 0.5 - 2.5 GPa 
: in this region, both SG and F phases coexist and their relative
amounts change with pressure.

  Initial measurements in the substituted
 series\cite{Katsufuji00} (RR')$_2$Mo$_2$O$_7$
 support a unique threshold from a ferromagnetic metal to an
 insulating spin glass state for a lattice constant {\itshape a$_c$}=10.33
\AA. Using the equation of state found above, it yields a chemical
critical pressure of 0.7 GPa. Actually, macroscopic measurements
on Gd$_2$Mo$_2$O$_7$ under pressure show a more complex situation.
Magnetic measurements under pressure \cite{Park03,Hanasaki06}
suggest a transition from ferromagnetic to spin glass state
already at 0.6 GPa. On the other hand resistivity
measurements\cite{Hanasaki06} on   {\itshape insulating}
Gd$_2$Mo$_2$O$_7$ crystals, show that they become metallic at a
much higher pressure of 2.4 GPa. It suggests that the
ferromagnetic-spin glass transition could be disconnected from the
metal-insulating one. We notice that all threshold values belong
the pressure range of coexistence deduced from our neutron data.


\begin{figure} [h]
\includegraphics* [width=\columnwidth] {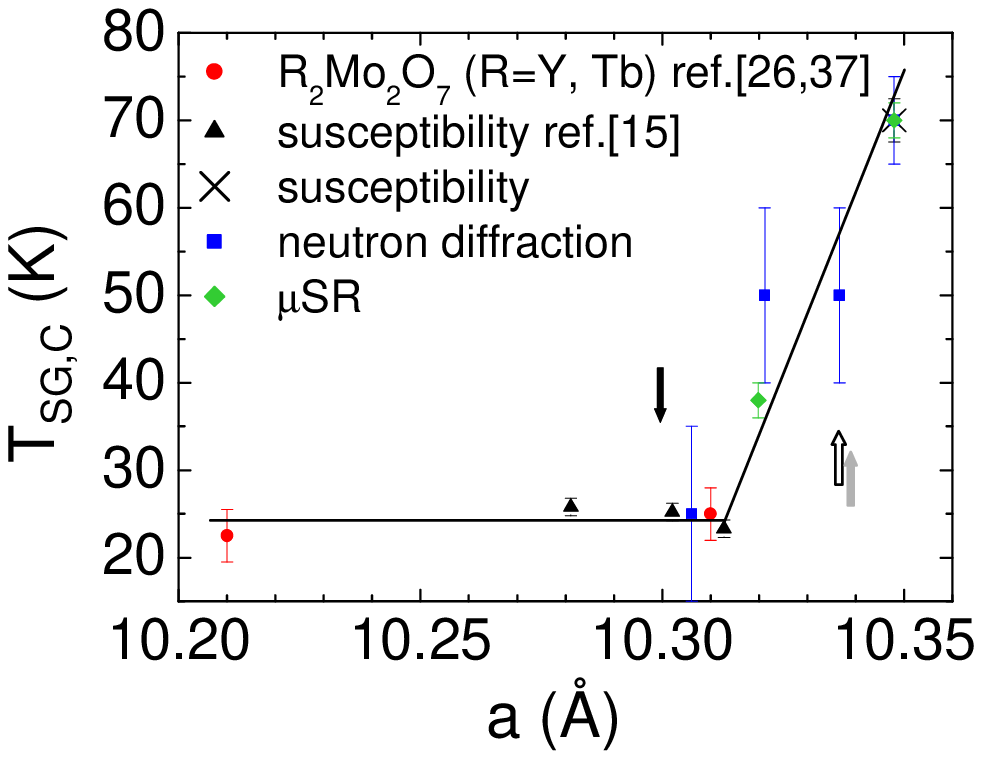}
\caption{Curie and spin glass transition temperatures T$_{\rm C}$
and T$_{\rm SG}$ versus lattice constant in
Gd$_2$Mo$_2$O$_7$:(crosses) from susceptibility, (squares) neutron
diffraction and (diamonds) $\mu$SR (this work); (triangles) from
susceptibility under pressure\cite{Hanasaki06}. The T$_{\rm SG}$
values of Y$_2$Mo$_2$O$_7$ ({\itshape a}= 10.21 \AA) and
Tb$_2$Mo$_2$O$_7$ ({\itshape a}= 10.31 \AA) from references
\cite{Gardner99,Gaulin92} are plotted for comparison. The arrows
indicate the critical lattice constant as determined from
susceptibility\cite{Park03} (grey), chemical
pressure\cite{Katsufuji00} (white) and
resistivity\cite{Hanasaki06} (black), using our compressibility
data.} \label{fig13.eps}
\end{figure}

In Fig. 13, we have plotted the transition temperature T$_{\rm C}$
measured in Gd$_2$Mo$_2$O$_7$ versus the lattice constant, by
combining our magnetization, neutron, muon and X ray data. We also
plotted the spin glass transition temperatures T$_{\rm SG}$
corresponding to the susceptibility anomalies under
pressure\cite{Hanasaki06}. These T$_{\rm SG}$ values are in the
temperature range 20-25 K, in good agreement with the values found
at ambient pressure in Tb$_2$Mo$_2$O$_7$ and Y$_2$Mo$_2$O$_7$ spin
glasses with smaller lattice constant\cite{Gaulin92,Gardner99}.
The temperature for the onset of short range correlations in the
spin glass state  is noticeably higher (around
30-40 K at 2.7 GPa) 
as it often occurs in spin glasses. The whole set of data provides
a precise description of the magnetic transition in the
instability region.

\begin{figure} [h]
\includegraphics* [width=\columnwidth] {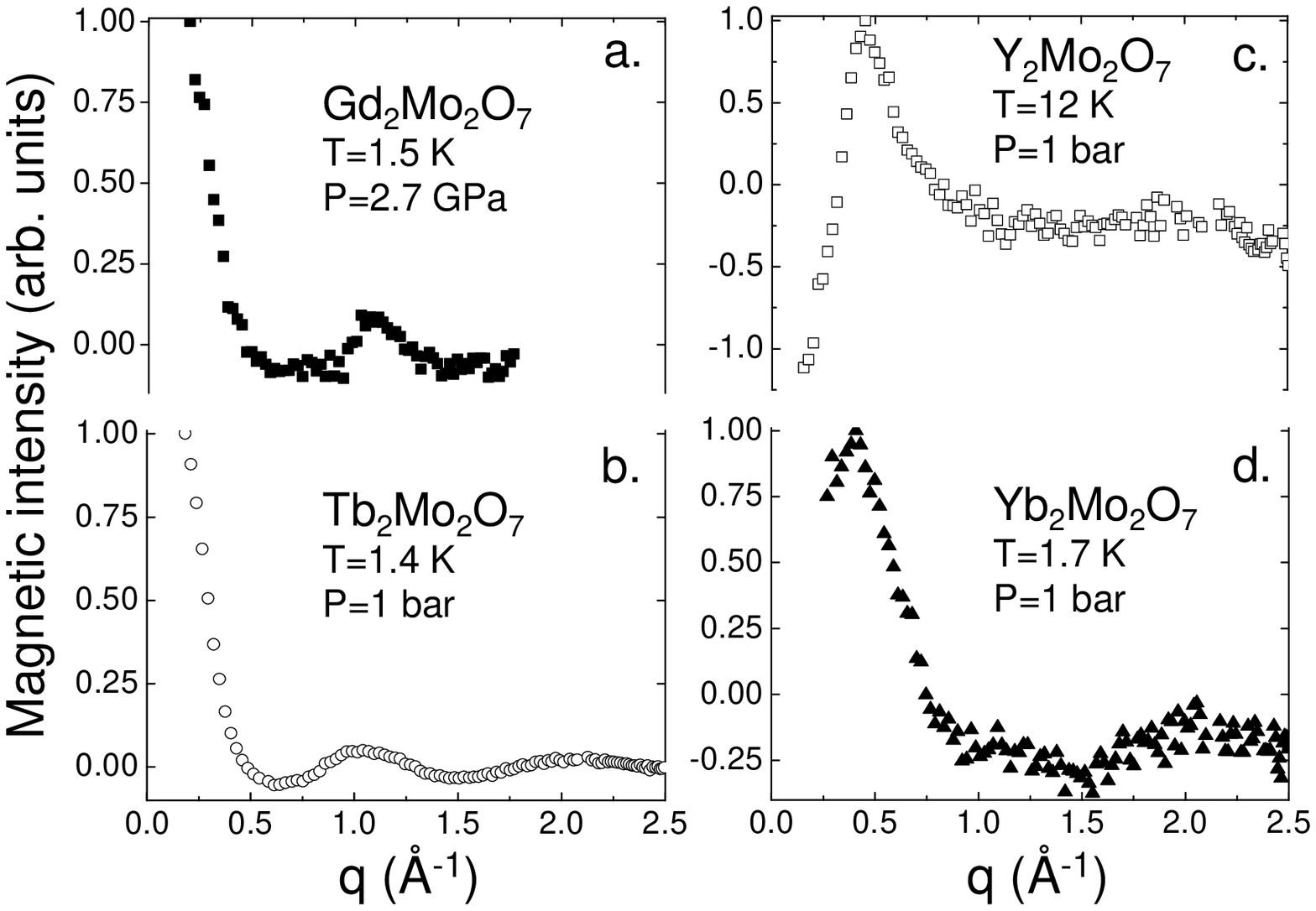}
\caption{Magnetic intensity of R$_2$Mo$_2$O$_7$ (R= Gd, Tb, Y and
Yb) versus the scattering vector {\itshape q}. 
A spectrum in the paramagnetic phase was subtracted. }
\label{fig14.eps}
\end{figure}

By comparing the magnetic pattern of Gd$_2$Mo$_2$O$_7$ at 2.7 GPa
with patterns of R$_2$Mo$_2$O$_7$ compounds with small ionic radii
(R=Tb, Y and Yb) at ambient pressure, we can get a general insight
on the different types of spin correlations in the spin glass phase.
For this purpose we measured the Y, Tb and Yb compounds at ambient
pressure (Fig. 14). Our results for R=Tb and Y agree with previous
results\cite{Gardner99,Greedan91,Gaulin92}. We can also follow the
evolution of the Mo-Mo, Gd-Mo and Gd-Gd correlations with pressure
from the ferromagnetic to the spin glass phase.

In the collinear ground state found in Gd$_2$Mo$_2$O$_7$ at
ambient pressure, all correlations are ferromagnetic. With
increasing pressure, Mo-Mo interactions become AF and frustrated
by the lattice geometry yielding the SG phase. Neutron scattering
cannot probe the Mo-Mo correlations in the SG phase of
Gd$_2$Mo$_2$O$_7$ since their contribution is much smaller than
the Gd-Gd and Gd-Mo ones due to the smaller Mo moment. But these
correlations can be directly evidenced in Y$_2$Mo$_2$O$_7$ where
only the Mo ions are magnetic and the lattice constant ( {\itshape
a}=10.21 \AA\ ) is equivalent to Gd$_2$Mo$_2$O$_7$ at 6.7 GPa. In
Y$_2$Mo$_2$O$_7$, AF correlations with a scale of about 5 \AA\,
forming a four sublattice structure\cite{Gardner99} yield a peak
in the diffuse scattering at {\itshape q}= 0.44 \AA$^{-1}$. A
similar structure is observed in Yb$_2$Mo$_2$O$_7$ ( {\itshape a}=
10.146 \AA\ ) where the contribution of the Yb$^{3+}$ moments
(around 1 $\mu_{\rm B}$) is much lower than for Tb and Gd.

Concomitantly with the change in the Mo-Mo correlations, the Gd-Mo
correlations become AF. The Gd-Gd correlations remain
ferromagnetic, but their correlation length decreases with
increasing pressure. Comparing Gd$_2$Mo$_2$O$_7$ with
Tb$_2$Mo$_2$O$_7$ suggests that ferromagnetic R-R correlations are
a general feature of the R$_2$Mo$_2$O$_7$ series. The length scale
of these correlations is reduced by pressure, which enhances the
AF interactions. Our measurements on Tb$_2$Mo$_2$O$_7$ under
pressure, to be published later, show that it is also the case for
Tb.

From our data, one can evaluate the first neighbor exchange
interactions ${\cal J}$ and their contribution to the energy
balance. One naturally expects ${\cal J}$$_{\rm Mo-Mo}$ $>$ ${\cal
J}$$_{\rm Gd-Mo}$ $>$ ${\cal J}$$_{\rm Gd-Gd}$. At ambient
pressure, the three interactions are likely
all ferromagnetic. 
Dominant Mo-Mo interactions stabilize the ferromagnetic order but
may be sensitive to temperature due to the band structure
instability. Gd-Mo ferromagnetic interactions account for the F
alignment of Gd and Mo moments.
 Gd-Gd interactions cannot be
 measured. One can speculate that their ferromagnetic character stabilizes the
 collinear state at low temperature.

We first estimate the exchange energy per ion (Mo or Gd) for the
three interactions. For Mo-Mo exchange, one can state that E$_{\rm
Mo-Mo}$ $\sim $ T$_{\rm C}$ $\sim $ $\theta_{\rm CW}$ = -70 K
where $\theta_{\rm CW}$ is the Curie-Weiss constant. For Gd-Mo
exchange, we use the exchange field acting on Gd, determined by
fitting the temperature dependence of the ordered Gd moment (Fig.
5): H$_{\rm ex}$ = 10.8 T; then: E$_{\rm Gd-Mo}$= -M$_{Gd}$H$_{\rm
ex}$ $\sim $ -45 K. Finally, to estimate the Gd-Gd exchange/dipole
energy, we use the $\theta_{\rm CW}$ value in Gd$_2$Ti$_2$O$_7$ or
Gd$_2$Sn$_2$O$_7$, where there is no 3d/4d magnetic metal ion:
E$_{\rm Gd-Gd}$ $\sim $ -10 K. Thus Gd-Mo and Gd-Gd interactions
may play a role to determine the ground state in the ferromagnetic
phase and threshold region.

 In order to
estimate the exchange integrals, we make the following
assumptions: the exchange between a pair of ions A and B is
written E$_{\rm ex}$ = -${\cal J}$$_{\rm A-B}$ S$_{\rm A}$S$_{\rm
B}$, where S$_{\rm A}$ and S$_{\rm B}$ are the true spins of the A
and B ions. Then, the exchange energy per ion is E$_{\rm A-B}$=
-z${\cal J}$$_{\rm A-B}$ S$_{\rm A}$S$_{\rm B}$, where z is the
number of nearest neighbors (z=6 in the pyrochlore lattice for all
exchange bonds). Although an ionic description is not fully
appropriate for the Mo ion in Gd$_2$Mo$_2$O$_7$, we assume S=1 for
Mo, and for Gd$^{3+}$, we use S=7/2 as there is no crystal field
splitting. Then, we obtain the following exchange integrals:
 ${\cal J}$$_{\rm Mo-Mo}$$\sim $ 12 K, ${\cal J}$$_{\rm Gd-Mo}$ $\sim $ 2 K and ${\cal J}$$_{\rm Gd-Gd}$ $\sim $ 0.14 K.

Under applied pressure, we find that Mo-Mo exchange interaction
changes sign and become AF. Calculated phase diagram for given
values of the exchange interactions\cite{Greedan91}, show that AF
Mo-Mo interactions may stabilize
 a degenerated ground state whatever the sign of the Gd-Mo interactions.
 Mo-Mo interaction involves Mo t$_{2g}$ orbital, which are very
sensitive to the electron correlations. Pressure is expected to
increase the intra site electron correlations energy U. At ambient
pressure, our estimation of ${\cal J}$$_{\rm Mo-Mo}$(12K) has the
same order of magnitude as the theoretical estimation of about 5
meV (58K), obtained by Solovyev\cite{Solovyev03} in the mean field
Hartree-Fock approach, taking the on-site Coulomb interaction U
=2.5 eV between Mo(4d) electrons. The model 
predicts that ${\cal J}$$_{\rm Mo-Mo}$
  decreases with increasing U, and one could speculate
 that in Gd$_2$Mo$_2$O$_7$, the ${\cal J}$$_{\rm Mo-Mo}$ exchange interaction becomes AF for a U value of about 3.5 eV,
 as it occurs in 
Y$_2$Mo$_2$O$_7$ for a lower U value. Gd-Gd interactions between
localized Gd moments are expected to be less sensitive to
pressure.

 The change the Mo-Mo interaction is connected with the aperture of
a Mott-Hubbard gap in the Mo t$_{2g}$ band, so that the
theory\cite{Solovyev03} predicts an insulating spin glass state.
The is actually true for the spin glasses with small ionic radii
(R=Y, Ho, Dy) but not for the pressure induced spin glass state.
Resistivity measurements\cite{Hanasaki06,Miyoshi03} on
Gd$_2$Mo$_2$O$_7$ and Sm$_2$Mo$_2$O$_7$ show that the spin glass
state induced under pressure is actually metallic. With respect to
chemical pressure, an applied pressure should not only increase
the electron correlations responsible for the Mott-Hubbard gap,
but also increases the band width. This should naturally favor
electron delocalization as a dominant feature. Optical
measurements could also check the evolution of conduction
properties under pressure.

 \section{Conclusion}

We have observed an abnormal ferromagnetic phase with strong
fluctuations in Gd$_2$Mo$_2$O$_7$. By applying pressure, we can
tune the change of this ferromagnetic phase into a spin glass
phase. The fact that an applied pressure is equivalent to a
chemical pressure to induce the ferromagnetic -spin glass
transition supports a mechanism mostly controlled by Mo-Mo
interactions. The combination of three microscopic probes under
pressure allows us to follow the evolution of magnetism with the
lattice constant with great precision throughout the threshold and
to evaluate the role of the rare earth interactions in the energy
balance.

We thank U. Zimmerman for the $\mu$SR measurements on GPD (PSI)
and M. Mezouar for the X ray measurements on ID27 (ESRF). We also
thank A. Forget and D. Colson for the sample preparation and
characterization. We are indebted to J. Rodr\'{\i}guez-Carvajal
for many useful discussions and for providing programs to analyze
the neutron diffraction data.

\end{document}